# Revealing the Unseen: The Discovery of Long-Awaited Radiation from the Intermittent Pulsar PSR B1931+24

Abdujappar Rusul[1], Z. G. Wen [2], J. P. Yuan[2], Ali Esamdin[2], X. P. Zheng[3], Michael Kramer[45]

**Affiliations:**

[1] Institute of Physics and Electrical Engineering, Kashi University, Kashgar, 844009, China.

[2] Xinjiang Astronomical Observatory, Chinese Academy of Sciences, Urumqi, 830011, China.

[3] Institute of Astrophysics, Central China Normal University, Wuhan, 430079, China.

[4] Jodrell Bank Centre for Astrophysics, School of Physics and Astronomy, University of Manchester, Manchester, UK.

[5] Max-Planck Institut für Radioastronomie, Bonn, Germany

Email: aai-pulsar@qq.com, yuanjp@xao.ac.cn , aliyi@xao.ac.cn, zhxp@mail.ccnu.edu.cn , michael@mpifr-bonn.mpg.de

**Abstract:** Pulsars are typically characterized by their stable, highly magnetized, and fast-rotating nature, which underpins their persistent radio emissions. However, the discovery of prolonged radio-quiet ("off") states in intermittent pulsars, such as PSR B1931+24, has challenged the most fundamental theory of pulsar magnetospheric emission. Despite long-term monitoring with several telescopes, continuous emission during these "off" states had not been detected in 20 years of observations. Fortunately, sensitive observations via Five-hundred-meter Aperture Spherical radio Telescope (FAST) revealed the mysterious weak emission containing occasional bursting dwarf pulses during the "off" states of PSR B1931+24. Along with a substantial decrease in flux density, a significant contraction in the integrated pulse width is measured in the "off" state compared to the radio-loud ("on") state, indicating alterations in the plasma supply and magnetospheric structure. Additionally, a previously unobserved dyssynchronous, nonuniform emission pattern is found in both states, supporting theories of a spatially inhomogeneous pair-cascade mechanism and challenging models of spatially coherent discharge. Furthermore, occasional dwarf pulses detected during the "off" state show flux and width distributions similar to those of the "on" state pulses, suggesting a potential link between the "on" and "off" state emissions of PSR B1931+24. Consequently, dwarf pulses are unlikely to represent a distinct emission mode as previously thought; instead, they appear to be part of a continuum within the pulsar's emission behaviors observed during the "on" state. These findings strongly support the basic theory of the pulsar magnetospheric emission and significantly advance our understanding of pulsar magnetospheric dynamics and their emission mechanisms.

**Main Text:** Radio observations of pulsars reveal a variety of emission properties, including periodic variations in profile shape and/or intensity (mode changing) (*1-3*), modulation of sub-pulses (sub-pulse drifting), short-term emission cessation (nulling) (*4-6*), long-term emission turnoff (intermittency) (*7*), etc. Mode changing may reflect underlying alterations in the pulsar magnetosphere's structure and emission characteristics (*3, 8, 9*). Sub-pulse drifting is often linked to the $\boldsymbol{E} \times \boldsymbol{B}$ drift of the charged plasma in the polar cap (*10*). Nulling is interpreted as failures in the intrinsic emission mechanism of pulsars (5, 7) or it could be attributed to cases such as the "breathing" of the magnetosphere, axion cloud dynamics in the polar cap, or short-term changes in the pulsar magnetosphere (*6, 8, 11, 12*). However, the intermittency of the emission in intermittent pulsars remains a puzzle since its discovery (*7, 13*).

To data, several intermittent pulsars have been discovered, including PSRs B1931+24, J1832+0029, J1841−0500, J1910+0517, and J1929+1357 (*7, 14-16*). Observations indicate that all the intermittent pulsars tend to slow down more rapidly during their "on" state compared to the "off" state, suggesting that the outflowing relativistic plasma associated with coherent radio emission in the "on" state significantly contributes to their spin-down (*7, 14, 15*). No associated X-ray or $\gamma$-ray emissions or glitches have been reported to explain the spin-down and emission feature of these pulsars (*7, 14, 15, 17*). Although a single short-period weak emission has been identified in PSR J1832+0029 during its "off" state, the most significant challenge posed by intermittent pulsars remains the prolonged cessation of emission lasting from months to years. This phenomenon cannot be explained by the basic theory of pulsar magnetospheric emission, which holds that the relatively stable magnetic field and rapid rotation of radio pulsars sustain their plasma-filled magnetosphere and its emission (*13, 18*). To explore the mystery surrounding the emission of the intermittent pulsar, this study focuses on the classic intermittent pulsar PSR B1931+24.

## A continues weak emission from the "off" state of PSR B1931+24

The about one-month period of the "off" state of PSR B1931+24 was described as complete emission cessation until recently, while the "on" state has the normal emission for about a week (*7, 19*). The super sensitivity of the FAST (*20*) enables us to detect the weak emissions, including occasional bursts of dwarf pulses, during the "off" state of PSR B1931+24, utilizing both the archival and the new observational data. Here is the pair of single pulse time series of the "off" state (at 58790 MJD) and "on" state (at 59131 MJD) data in Fig. 1. The data is calibrated by using the radiometer equation $S_{av} = S_{sys}/\sqrt{N_p F_e T_{int}}$ (*21-23*), where $S_{sys} = 1$ Jy for FAST (*24*), $N_p$=4 is the number of polarizations summed, $F_e$=500 MHz is the effective bandwidth, $T_{int} = 256\tau_s$ is integrated time of a single data point, the sampling time ($\tau_s$) is 49.152 µs, then the off-pulse noise $S_{av}$= 0.2 mJy, and the flux density can be obtained by $S_{peak} = S_{av} S/N$. As shown in Fig. 1, the pulse flux density during the "off" state decreases dramatically compared to the "on" state. Single pulse analysis detected a number dwarf pulses during the each "off" state, with peak signal-to-noise ratios (S/N) ranging from 5 to 30 (Table S1), which corresponds to a flux density of 1 to 6 mJy. They predominantly display a sporadic, narrow burst pulse pattern during the "off" state, in contrast to the normal emission observed in the "on" state (illustrated in the insets of Fig. 1). The occurrence rate of individual dwarf pulses during the "off" state is approximately 1% (Table S1).

Due to the spin-down rate transition of the intermittent pulsars, the phase connection of the different observing data of PSR B1931+24 cannot be done yet. The data in the "off" state being folded according to the pulse period obtained from the "on" state data. Based on the integrated pulse profiles of the each observing data, the pulse longitude ranges for the pulse-on window are selected. The mean flux density of the observed data for both the "on" and "off" states are then evaluated (the 7-8th columns of Table S1 and Fig. S2), in which only the individual bursting dwarfs are considered for the "off" state. The average flux density of the pulse-on window of the individual pulses in "on" and the individua dwarfs in "off" states are $S_{mean}^{on} = 120.78 \pm 15$ mJy and $S_{mean}^{off(d)} = 5.70 \pm 0.8$ mJy, respectively, i.e., $(S_{mean}^{on})/(S_{mean}^{off(d)}) \approx 21$. It can be also noted in columns 7–9 of Table S1 that, for data collected between MJDs 60561 and 60596— which have nearly identical observing durations—the average ratio of the peak-to-peak S/N of the integrated pulse during the "on" states to that of the total emission in the "off" states is approximately 285. This implies the significant decrease in intensity of the emission during the "off" state compared to the "on" state.

Besides the detection of significant weak emission in the form of total or bursting dwarf pulses,

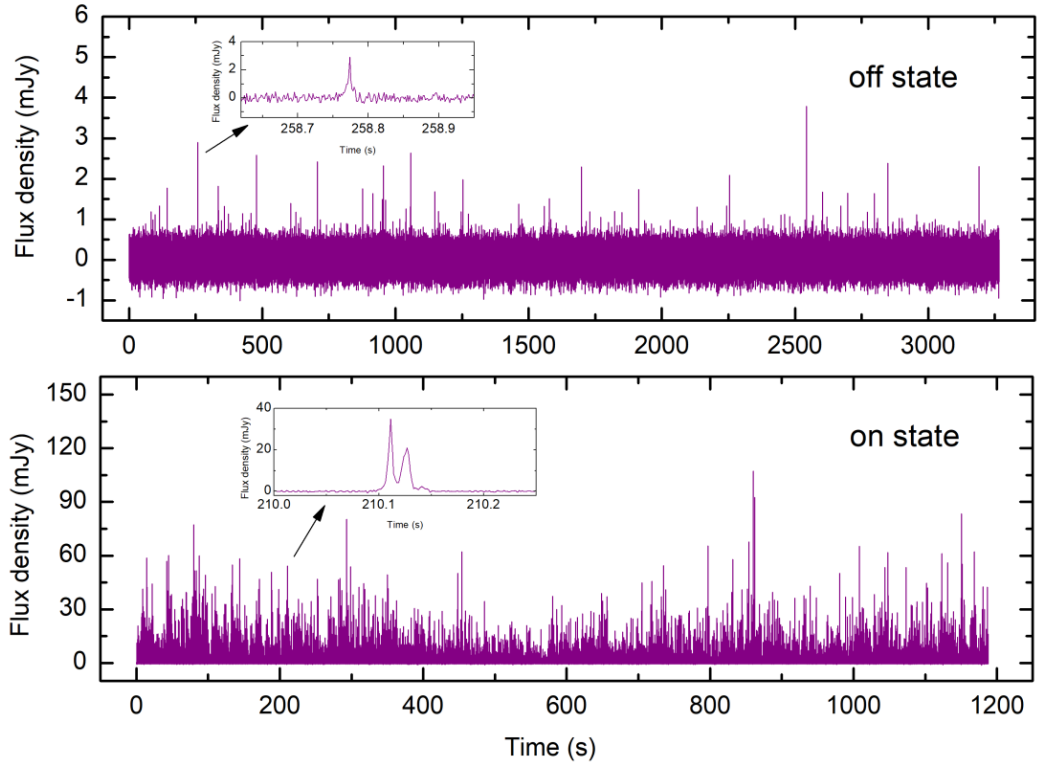

Fig. 1. The single pulse time series of PSR B1931+24. The upper plot is the "off" state at 58790 MJD; the lower plot is the "on" state at 59131 MJD, respectively (the main plot). The inset in the upper plot is the one of the individual bursting dwarf pulses in the "off" state; the inset in the lower plot is the one of the regular individual pulses in the "on" state.

the analysis of the total minus bursting dwarfs (total–dwarf) also reveals notable emission during the "off" states—excluding the data at 59194 MJD, which has the shortest observation time compared to other "off" state data—with peak S/N values ranging from 5 to 10 and flux densities approximately 1 to 2 mJy (see Table S1). This indicates the presence of even weaker emissions beyond the bursting dwarfs during the "off" state of PSR B1931+24. With increased observational time and sensitivity, these underlying weak emissions could be detected, suggesting that there may no longer be a complete, prolonged emission cessation during the "off" state of this intermittent pulsar.

## The changing pulse profiles

The data presented in this study include two "on" states, seven "off" states, and one transition state from "off" to "on" (see Table S1). It is observed that the shapes of the integrated pulse profiles in the two "on" states and the transition state are nearly identical. Here, for more accurate measurement of the pulse width, the transition state (at 60561 MJD) is selected, as it features a relatively longer duration of "on"-state emission compared to the other two "on" states. The full width at half maximum (FWHM) of the integrated pulse profile, $w_{50}$, is measured to be 10.11° (Table S2). However, this result is slightly smaller than the value of 11.46° reported on the Pulsar Catalog Database[1] (25). This discrepancy is likely due to the shorter duration of our "on" state data compared to the observations from the Lovell and Mark II Telescopes at Jodrell Bank (19).

Regarding the "off" state emission, the longest single observing session at MJD 58790 is preferred. The integrated pulses of total, dwarfs, and total-dwarfs (the weaker pulses, other than the dwarf pulses) are normalized by their own peak value (see the left plot in Fig. 2); there is a notable difference in the pulse profile shape between "on" and "off" state. The $w_{50}$ of the integrated pulse profile for the total is 7.57°, indicating that the "off" state pulse profile width shrinks by up to 34% compared to the corresponding "on" state value (11.46°) reported in the Pulsar Catalog Database (25), or by about 25% based on the data from FAST (Table S2). These variations in pulse flux, profile width, and shape, along with the change in spin-down rate between the "on" and "off" states, imply substantial changes in the structure and plasma supply of PSR B1931+24 during its state-switch (8, 26).

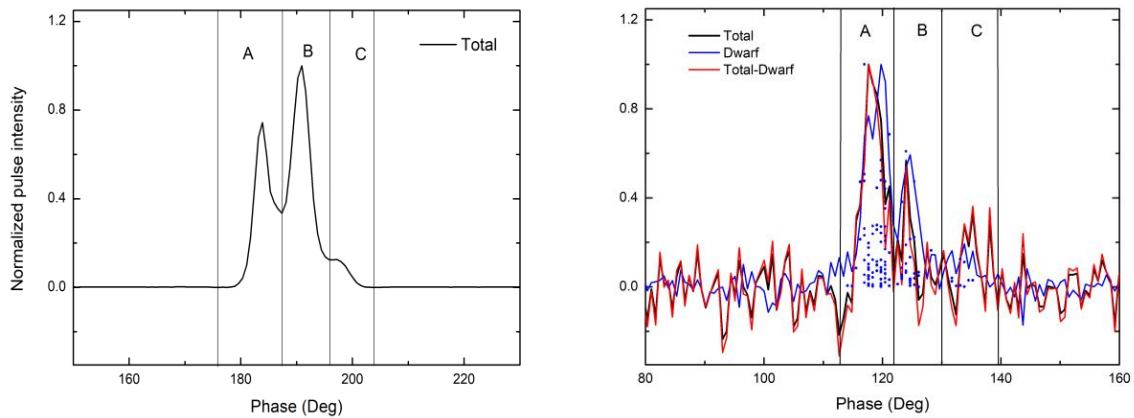

Fig. 2. The integrated pulse profile of PSR B1931+24 in its "on" and "off" state. The integrated pulse profile of total is on the left, with data at 60561 MJD; and the profiles on the right are the integrated pulse profiles of total (thick black), the dwarfs (blue), and the total-dwarf (red), respectively. The blue dots are the single pulses with a S/N>5, which constitutes the bursting dwarfs.

As shown in Fig. 2, the integrated pulse profiles in both "on" and "off" state exhibit a multi-peak structure (see the left plot in Fig. 2) and can be described as consisting of three main components: the leading, central, and trailing components, which are labeled as A, B, and C. To explore the emission feature, we analyzed the intensity distribution of the total pulse, as well as components A, B, and C (Fig. 3S). In "on" state: the distribution of the total intensity is clearly separated from the off-pulse noise, indicating the absence of nulling in the "on" state. It is also observed that the intensity distributions of the total pulse, as well as components A and B, exhibit similar characteristics and can be well-fitted using a lognormal distribution model. In contrast, the distribution of component C partially overlaps with the off-pulse noise, which follows a Gaussian shape. This suggests that the emission of component C has an intermittent feature. In "off" state: the intensity distributions of the total, A, B, and C overlap significantly with off-pulse noise, suggesting an increased number of nullings in the form of entire individual pulses and/or nullings localized to their subcomponents. However, the integrated pulse profile of the total-dwarf (shown as the red line on the right plot of Fig. 2) still exhibits notable weak emission, with a S/N of 8 at its peak (see Table S1), and its pulse structure closely resembles that of the total pulse profile (the peak S/N of 10). It is noticed that the potential weaker emissions beyond the dwarf pulses become significant after integrating approximately one thousand individual pulses. Consequently, pulsars that display dwarf pulses during their nulling states may also emit weak signals that could be detectable with sufficient observing (or integration) time and sensitivity (*27, 28*).

---

[1] atnf.csiro.au/research/pulsar/psrcat/. The results for PSR B1931+24 in the Pulsar Catalog Database are based on observations at center frequencies ranging from approximately 1350 to 1500 MHz, which are close to FAST's observing frequency of 1250 MHz. Therefore, the influence of the observing frequency on the measurement of pulse width can be considered negligible.

In the "off" state at MJD 58790, 63 individual dwarf pulses were identified among a total of 4,055 individual pulses (see Table S1), and they exhibit a quasi-periodic nature (see Fig. 4S). The bursting dwarf pulses (depicted as blue dots in the upper left plot of Fig. 2) are distributed throughout the entire on-pulse window but are predominantly concentrated around the leading component A. In contrast, the dwarf pulses from PSR B2111+46 are primarily distributed around the trailing component (*27*). Considering a core and conal polar cap emission structure and the asysmmetric polar cap current density distribution (*21, 29, 30*), these dwarf pulses may be associated with emission processes near the last open field lines of the pulsar's magnetosphere. In this region, the necessary charges for polar cap emission might be supplied via reconnection near the Y-point — the boundary between closed and open magnetic field lines near the equatorial plane — through the separatrix layer, which separates closed and open field line regions of pulsar (*26, 31*).

## Dyssynchronous nonuniform emission

The component nulling feature of the individual pulses implies a nonuniform dyssynchronous emission across the magnetic field line of the pulsar. To further investigate the emission feature of PSR B1931+24, the individual pulses in both states are reorganized basing on the intensity of A, B, and C, and total (Fig. 3). Notably, comparing with original pulse stacks given in Fig. S5, the components of the individual pulses in both states clearly display mutually constraining patchy emission patterns that they manifest a kind of inverse features, e.g., when component A becomes brighter, component B gets dimmer, and component C lessens or disappears, etc. Comparable emission patterns are also observed in the remaining datasets, though they exhibit temporal variability (Fig. S6). This phenomenon has not been explored in any other pulsars.

If the observed dyssynchronous striped patchy emission pattern can be associated with the bunching discharge spot in the polar cap, it suggests the presence of spatially dyssynchronous, long-lasting intermittent discharges that persist along the same field lines for at least one on-pulse window period of pulsar rotation. This strongly supports the existence of the spatially inhomogeneous pair-cascade mechanism proposed in recent theoretical work (*32*). Conversely, these findings challenge the current pulsar polar cap pair-cascade models, which predict spatially coherent discharge dynamics (*6*).

For the quantitative analysis of putatively associated pulse components A, B, and C, we adopt the average intensity of each component as a reference baseline. A decrease in intensity is defined as any component falling below its baseline value, whereas an increase corresponds to exceeding this threshold. Analysis of data from MJDs 58790 and 59131 reveals that 28% (on-state) and 39% (off-state) of cases exhibit synchronous intensity decreases/increases across all three components, while 72% (on-state) and 61% (off-state) display asynchronous behavior (i.e., inverse relationships between components). Similar trends are observed across other datasets in Table S1, with average synchronous changes occurring in 27% (on-state) and 44% (off-state) of cases, contrasted by asynchronous variations in 73% (on-state) and 56% (off-state). It is worth noting that when considering the brighter individual pulses in the upper sections of the plot in Fig. 3, the asynchronous feature can reach up to 90% in both states. This suggests a potential link between pulse intensity and the observed dyssynchronous emission, as well as the corresponding asynchronous pair-cascade. These findings indicate significant spatial incoherence among the components in both states. However, the physical mechanisms driving these variations remain unclear and warrant further investigation.

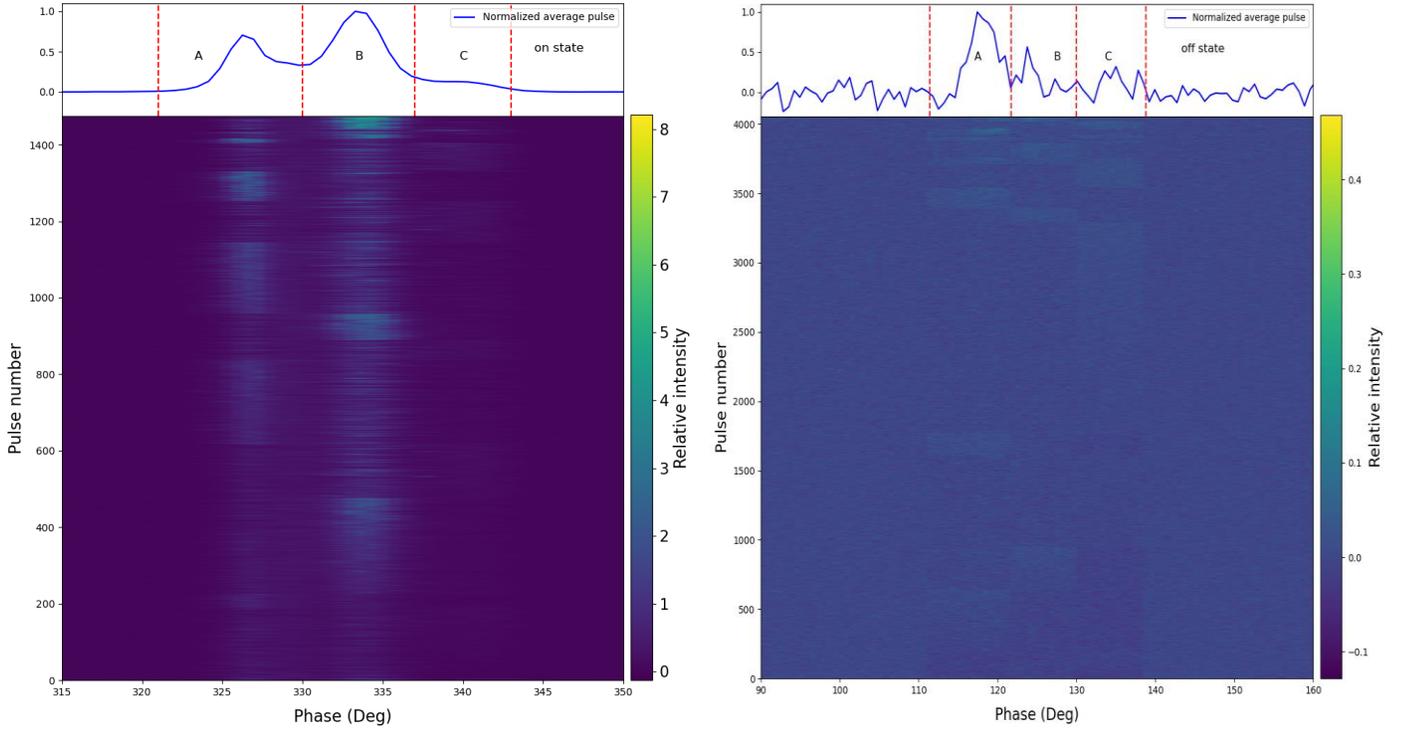

Fig. 3. Reorganized pulse stacks for the "on" (left) and "off" (right) state data recorded at 59131 and 58790 MJDs, respectively. A distinct dyssynchronous, striped patchy emission pattern is evident among components A, B, and C of the individual pulses.

**Dwarf pulse emission**

For the first time, the observation at 60561 MJD of PSR B1931+24 captured a state transition from "off" to "on" within one rotational period, in which the intensity and the pulse structure demonstrate a gradual recovery feature (see Fig. 4 and Fig. S7). Single-pulse analysis and integrated pulse profile reveal no significant emission during the initial 13 individual pulses, followed by the abrupt emergence of emission in the 14th individual pulse, as illustrated in the right and left insets of Fig. 4. This 14th individual pulse marks the onset of the state transition, exhibiting a peak S/N of 13 (corresponding to a flux density of 2.6 mJy) and a pulse width ($w_{50}$) of 1.80°, manifesting typical dwarf pulse feature. The triggering mechanism of this dwarf pulses may be the source of state-switching of the PSR B1931+24.

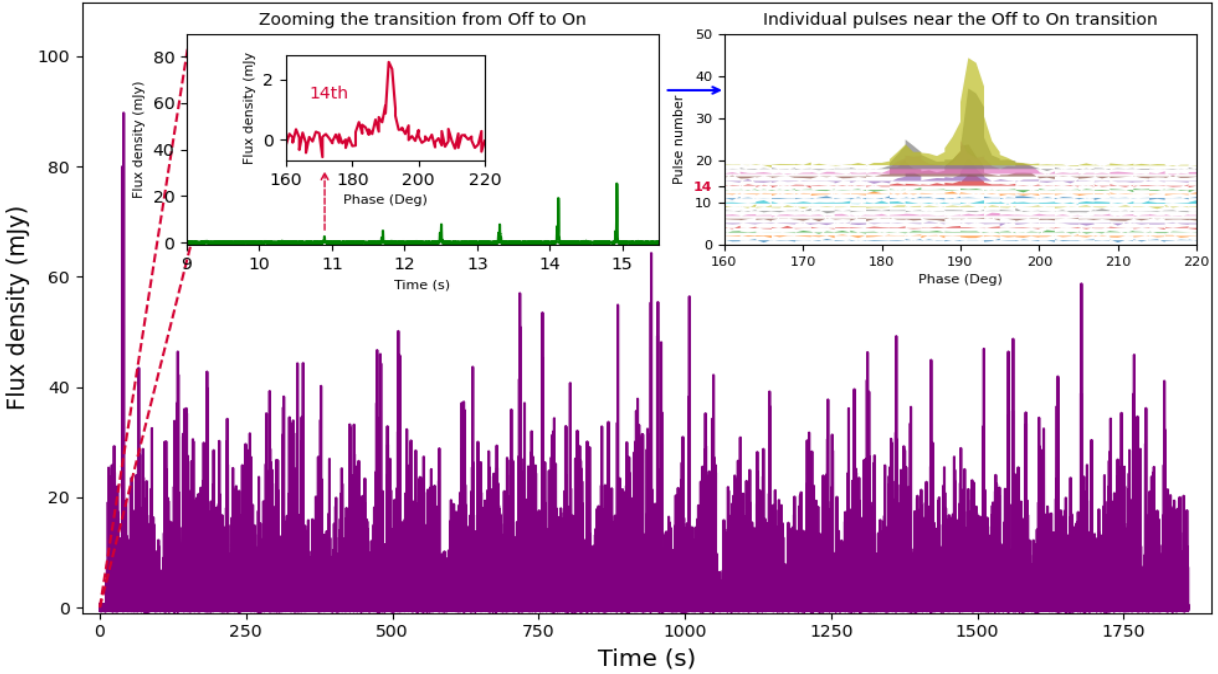

Fig. 4. The single pulse time series of PSR B1931+24 shows an off-to-on state transition at 60561 MJD (main plot). The transition period is highlighted in the upper left inset (green), with the start of the transition from "off" to "on" indicated in the 14th individual pulse (red). The upper right inset displays the individual pulses before and after the transition.

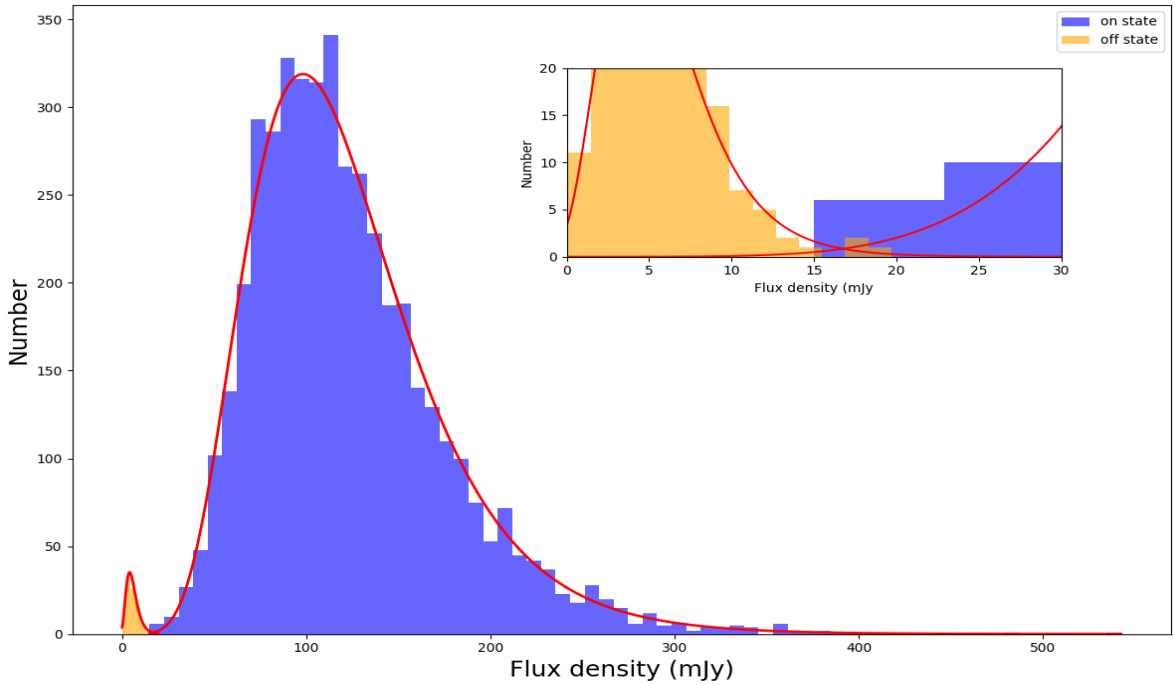

Fig. 5. The flux density distribution of the individual pulses in "on" states and bursting individual dwarf pulses in "off" states. They both follow a lognormal distribution (red line). The inset shows the overlapping region of the flux distribution between "on" and "off" states.

The previous studies did not find the distribution feature of the dwarf pulses (*27, 28*), which would provide the underlying relationship between the two different emission state of the pulsar. As shown in Fig. 5, the data of PSR B1931+24 revealed that the pulse flux density distribution of the "on" state individual pulses and "off" state bursting individual dwarf pulses both follow lognormal distributions, and they overlapped with each other at their margins (see Fig. 5). These imply that the emion in "on" and "off" state is not completetely different, but a similar emission mechanism might undergo during the "on" and "off" states.

To further explore the relationship between the "on" and "off" states, we analyze the pulse flux density distribution relative to the pulse width ($w_{75}$) for all "on"-state individual pulses and "off"-state bursting individual dwarf pulses. The "on"-state data exhibit a bimodal distribution (red, blue, and green points in Fig. 5), possibly suggesting the presence of two distinct radiation processes within the pulsar magnetosphere. Meanwhile, the "off"-state dwarf pulses predominantly cluster beneath the narrower, brighter "on"-state pulses in this parameter space. This alignment may also indicate a shared emission mechanism between the bursting dwarf pulses and their brighter "on" state counterparts.

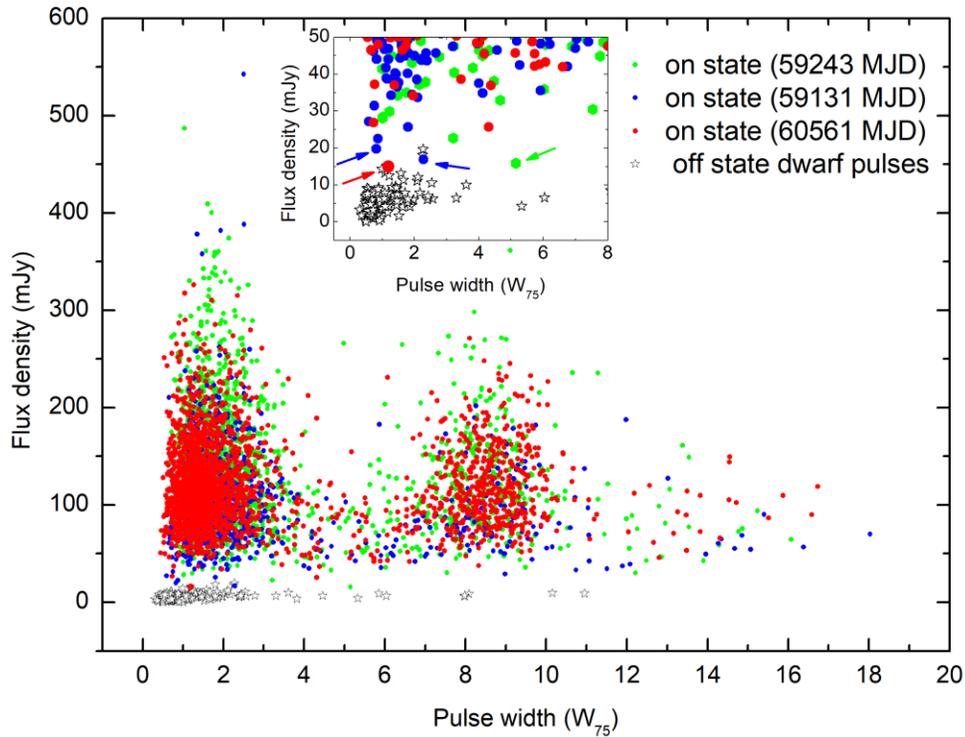

Fig. 6. Pulse flux density versus pulse width ($w_{75}$) distributions for individual pulses in the "on" (colored dots) and "off" (black hollow stars) states. Vertical axis represents the summed flux density within the pulse window, defined by the integrated pulse profile. The pulse width ($w_{75}$) of the individual pulse is estimated at 75% of its peak due to some of the $w_{50}$ of the individual pulses in the "off" state being contaminated by the off-pulse noise. The red, blue, and green dots in the plot are from the data in the "on" state of 60561, 59131, and 59243 MJDs, the black hallow stars at the bottom left corner of the plot are from the individual dwarf pulses in all of the "off" state data. The inset highlights that the 14th individual pulse from the transition state (shown in Fig. 4) is located within the region where dwarf pulses are distributed (the red dot indicated by the red arrow in Fig. 5). Additionally, one green dot and two blue dots from the "on" state data at 59131 and 59243 MJDs also reside within this region.

In addition, an interesting finding is that the 14th transition-state individual pulse (indicated by the red arrow in Fig. 6) and three "on"-state individual pulses (one green from 59243 MJD and two blue from 59131 MJD, indicated by green and blue arrows, respectively) reside within the dwarf pulse region (see Fig. 6). This positioning illustrates again the overlapping distribution of "on" and "off" state emissions at their margins (see also Fig. 5). This further indicates that the bursting dwarfs in the "off" state and the brighter narrow "on"-state emission might have similar magnetospheric origin, implying that dwarf pulses unlikely constitute a separate emission mode as previously described [11], but rather part of a continuum within pulsar emission behaviors.

## Possible dynamics of the pulsar magnetosphere

The cessation of pulsar radio emission is theorized to arise from the collapse of charge acceleration mechanisms within key magnetospheric regions, such as the polar cap (*10*), slot gap (*33*), outer gaps (*34*) in the open field line zones, and the Y-point region (*35*). The radio emission features of PSR B1931+24 can be associated with polar cap dynamics (*7, 21*). Based on the polar cap emission theory, the emission suppression occurs when dense plasma, through a discharging process above the polar cap, screens the polar gap's acceleration field—a transient process lasting mere fractions of a spin period (*30, 36*). This raises a critical question: How can intermittent pulsars cease their emission for such extended periods while exhibiting normal emission like other pulsars during their "on" state? If we lack observational evidence or theoretical frameworks to suggest drastic changes in fundamental pulsar parameters—such as magnetic field strength and/or spin period—to explain the exotic emission feature of the intermittent pulsar, the most plausible explanation is that weak emission may still occur during the "off" state but its intensity is too low to be detected by most of the currently operating radio telescopes or it needs to refine the fundamental model of pulsar magnetospheric emission. In this context, the FAST detection of weak emission from the "off" state of PSR B1931+24 strongly supports the notion that pulsars cannot cease emission in a prolonged manner unless they genuinely die.

Within the magnetic dipole model of pulsar magnetospheres, the extent of the closed field line region and polar cap opening angle are associated with the pulsar's spin-down rate as $\dot{v} \propto \left(\frac{1}{r_{open}}\right)^2 \propto (\theta_{pc})^4$ (*8, 26, 37-39*), where $\dot{v}$ is the spin-down rate of pulsar rotation, $r_{open}$ is the extent of the corotation zone of the pulsar magnetosphere (i.e., the closed field line region), $\theta_{pc}$ is the polar cap opening angle. In this case, from the change in spin-down rate, (i.e., $|\dot{v}_{on}| > |\dot{v}_{off}|$ and $\frac{\dot{v}_{on}}{\dot{v}_{off}} = 1.5$ or $\frac{\Delta\dot{v}}{\dot{v}_{on}} = 0.34$ (*7, 19*)), of PSR B1931+24 in between the "on" and "off" states, one can find 22% ($\frac{\Delta r_{open}}{r_{open(on)}} = 0.22$) expansion of the closed field line region, and corresponding 11% ($\frac{\Delta \theta_{pc}}{\theta_{pc}^{off}} = 0.11$) contraction of the polar cap emission cone. According to the polar cap emission theory[2], the "on"-state pulse profile of PSR B1931+24 possibly suggest a core-cone structure in its polar cap emission region; if considering the impact parameter $\beta \sim 0$ for the seemingly central traverse of our line of sight on the polar cap emission cone of PSR B1931+24 (*8*) and $w_{10}^{on}/4 < 5°$ for the "on" state, the observed 34% (or 25%) reduction in pulse width ($w_{50}^{on}$) during the "off" state implies an approximately proportional decrease in the polar cap opening angle, i.e., $\frac{\Delta w}{w_{on}} \approx \frac{\Delta \theta_{pc}}{\theta_{pc}^{on}} \approx$ 0.34 (or 0.25). Alternatively, the change in the spin-down rate of intermittent pulsars could be explained by variations in the plasma supply within the polar cap or across the entire open field line region of pulsar magnetosphere (*7, 40*). However, the huge vacuum potential in the open field line region is hardly consistent with the prolonged "off" state and/or the observed weak emissions of PSR B1931+24 in this study.

---
[2] The polar cap emission geometry shows that $(\sin\frac{W}{4})^2 = \frac{(\sin\frac{\theta_{pc}}{2})^2 - (\sin\frac{\beta}{2})^2}{\sin\alpha \, \sin(\alpha+\beta)}$, where $w$ is the pulse width, $\alpha$ is the angle between the rotational axis of the pulsar and the line of sight (inclination angle), $\beta$ is the angle between the magnetic axis and the line of sight

Neither purely geometric models nor treatments based on plasma supply and current density adequately explain the simultaneous changes in spin-down rate, pulse width, and flux density observed in PSR B1931+24. A comprehensive model is necessary to elucidate the correlations among these changes between the "on" and the "off" states. From a physical perspective, the expansion of the closed field line region within the pulsar's magnetosphere would compress or suppress the original polar cap area (*8*). This would lead to a decrease in charged plasma density and accelerating potential, making it difficult to sustain the conditions necessary for pulsar radio emission. This deficit could potentially be compensated if sufficient charges are returned to the polar cap through a possible magnetic reconnection event at the Y-point region, which might sporadically trigger bursts of dwarf pulses and subsequent weak emissions. This mechanism offers a plausible explanation for PSR B1931+24's emission behavior in the "off" state. But why does the structure of the pulsar magnetosphere change for PSR B1931+24? and What causes the transitions between the "on" and "off" states?

Pulsars function as self-sustaining charge generators powered by their immense magnetic fields and rapid rotation, continuously producing the necessary charges through pair-production processes within their magnetospheres (*13, 21, 41, 42*). Meanwhile, charges in the open field line regions are either ejected as a pulsar wind or accumulate at the Y-point region. At this critical junction, magnetic reconnection events routinely redirect charges through the separatrix layer to replenish the acceleration field in the polar cap (*26, 31*). During a pulsar's regular emission phase, its magnetosphere would undergo transient charging through partial reconnection of the outermost open magnetic field lines. This process modestly expands the closed field line region, temporarily trapping additional plasma. However, when the expansion reaches to its limit, the polar cap may no longer maintains its actual shape and function, leading to diminished or ceased emission. Then, discharging may dominate at the Y-point region through the magnetic reconnection (i.e., the previously closed lines reopen), allowing some of the discharging charged plasma to return to the polar cap for triggering and/or enhancing the polar cap emission, while the rest moves outward to contribute to high energy emissions within the equatorial current sheet beyond the light cylinder of the pulsar magnetosphere. This discharging process would restore the magnetosphere to its original configuration and function over time.

Intermittent pulsars, which possibly have critical polar cap configurations—such as a gap height approximating the polar cap radius (*43, 44*) — would possess a fragile magnetospheric equilibrium. As described above, during the "on" emission phases (lasting weeks to years) of the intermittent

---

(impact parameter), and $\theta_{pc}$ is the opening angle of the polar cap (21). If considering smaller $w$ and $\theta_{pc}$, and $\beta \sim 0$ (or $\beta \ll \theta_{pc}$) for the seemingly central traverse of our line of sight on the polar cap emission cone of PSR B1931+24 (17), one can obtain $\frac{\Delta w}{w_{on}} \approx \frac{\Delta \theta_{pc}}{\theta_{pc}^{on}}$ for the "on" and "off" state of the intermittent pulsar. And due to $w \propto w_{10} \propto w_{50}$, we may have $\frac{\Delta w}{w_{on}} \sim \frac{\Delta w_{50}}{w_{50}^{on}} \sim \frac{\Delta \theta_{pc}}{\theta_{pc}^{on}}$.

pulsars (*7, 14, 15*), the closed magnetic field line zone of the magnetosphere is believed to gradually expand until it reaches a critical threshold that triggers a transition to the "off" state. Immediately after the state-switch, discharges at the Y-point region initiates the "off" state emission as seen in PSR B1931+24, and the pulsar reverts to its normal emission phase after some periods of magnetospheric evolution. In this context, the possible reconnection mechanism near the Y-point region of the pulsar magnetosphere provides a promising explanation for the state-switching mechanism of intermittent pulsars.

The weak emission from the "off" state of the classic intermittent pulsar PSR B1931+24 testifies to the basic theory of the pulsar magnetospheric emission. This indicates that the pulsar magnetosphere is not completely depleted of charged plasma even in the temporary "off" state; therefore, both the "on" and "off" states of PSR B1931+24 can be described by a force-free model, except for the regions of charge acceleration and pair cascades. The state-switching behavior of PSR B1931+24 likely results from a dynamical magnetospheric reconfiguration, which is driven by a self-regulating interplay between plasma dynamics, magnetic field topology, and pair production processes. Cooperative cadent observational campaigns with advanced facilities such as FAST, Lovell, and Mark II Telescopes at Jodrell Bank, etc., are needed to obtain the complete timing solutions of this pulsar, which has profound importance to carry out an in-depth study of the pulsar emission, magnetospheric dynamics, and spin-down mechanism.

**Acknowledgments:** This work has used the data from the Five-hundred-meter Aperture Spherical radio Telescope (FAST). FAST is a Chinese national mega-science facility, operated by the National Astronomical Observatories of Chinese Academy of Sciences (NAOC). We wish to thank an anonymous referee for helpful comments. This work was funded by the National Natural Science Foundation of China (NSFC) under grant No. 12103021, the Key Program of National Natural Foundation of China 12033001, the National SKA Program of China grant No. 2020SKA0120300. Some of the data plotting code used in this study was enhanced with assistance from the Sider AI platform, which provided valuable resources and examples.

**Data availability:**

All the data that support our results and analysis are available from the first author on reasonable request. The observational data from 58790 to 59243 MJDs are publicly available from the FAST archival data base (https://fast.bao.ac.cn/), and the newly observed data are open source after the one-year protection for the high-priority usage by observers, according to the FAST data policy.

**Supplementary Materials**

A simple description of the data is given in Table S1. Data is a PSRFITS format (*45*), and it is 8-bit sampling, the center frequency is 1250 MHz with 500 MHz bandwidth (*46, 47*). PSR B1931+24 has a period of 813.7 $ms$, and its dispersion measure (DM) is 106.03 $pc\ cm^{-3}$ (*48*). The integration of the individual pulses reveals weak emissions in the "off" state data of PSR B1931+24. To further confirm, DM of 0-150 $pc\ cm^{-3}$ ranges are tried by the Presto package (*49*) on the "on" state data from 58790 MJD and "off" state data from 59131 MJD. We observed that the faint pulse in the "off" state appears most prominent in the DM-time plot, with a DM value consistent with that in the "on" state (Fig. S1). Subsequently, we employed the Tempo2 (*50*) package to update the ephemeris of PSR B1931+24. Following this, the DSPSR (*51*) package was utilized with the new ephemeris on the raw data, reducing it to 512 bins (each bin containing the energy from 32 sampling points). We then used the PAZ and PAZI tools from the PSRCHIVE (*45*) software to mitigate radio frequency interference (RFI). Finally, the PSRCHIVE tool PAM was used to compute the total intensity (I) of the single pulses.

Table S1. Ten epochs of the observing data of PSR B1931+24

| MJD | Observation duration (s) | Individual pulses | Observation mode | State of emission | The number of individual Dwarfs / The occurring rate | The peak S/N of Total (integrated pulse)/ Mean flux density (mJy) | The peak S/N of Dwarfs (integrated pulse) /Mean flux density (mJy) | The peak S/N of Total - Dwarfs (integrated pulse) | The maximum S/N of single pulses |
|---|---|---|---|---|---|---|---|---|---|
| 58790 | 3600 | 4055 | Tracking | Off | 63/1.6% | 10 | 20/5.73 | 8 | 19 |
| 59131 | 1200 | 1475 | Tracking | On | 1* | 2341/135.70 | | | 538 |
| 59194 | 735 | 737 | Swift Calibration | Off | 8/1.1% | 4 | 16/5.52 | 3 | 30 |
| 59243 | 735 | 738 | Swift Calibration | On | 2* | 1501/105.13 | | | 504 |
| 60561 | 1940 | 2310 | Swift Calibration | Off-to-On | 1* | 2168/121.51 | | | 450 |
| 60568 | 1940 | 2310 | Swift Calibration | Off | 21/0.9% | 7 | 24/6.38 | 5 | 16 |
| 60575 | 1940 | 2310 | Swift Calibration | Off | 17/0.7% | 6 | 11/5.01 | 5 | 19 |
| 60582 | 1940 | 2310 | Swift Calibration | Off | 17/0.7% | 6 | 12/5.49 | 5 | 19 |
| 60589 | 1940 | 2310 | Swift Calibration | Off | 27/1.2% | 8 | 19/7.14 | 6 | 16 |
| 60596 | 2180 | 2604 | Swift Calibration | Off | 30/1.2% | 11 | 19/4.66 | 10 | 18 |

* Refers to the pulses in the "on" state which have a similar distribution feature as dwarf pulses in the "off" state, as shown in Fig. 6.

Table S2. Pulse profile widths of the integrated total pulses of PSR B1931+24 in its "on" and "off" states.

| Referenced data | $w_{10}^{on}$ (deg) | $w_{10}^{off}$ (deg) | $w_{40}^{on}$ (deg) | $w_{40}^{off}$ (deg) | $w_{50}^{on}$ (deg) | $w_{50}^{off}$ (deg) |
|---|---|---|---|---|---|---|
| FAST | 17.47 | | 10.71 | 8.12 | 10.11 | 7.57 |
| Pulsar Catalog Database | 18.10 | | | | 11.46 | |
| The change rate of pulse width $\Delta w/w_{on}$ (%) | | | | 24% | 25%/34% | |

Note. The "on" and "off" state pulse widths from the FATS are obtained by using the longest observing data at 60561 and 58790 MJDs, respectively. The pulse width change rate of 25% is measured by using the data from FAST. while the change rate of 34% is measured by using the $w_{50}^{on}$ from the Pulsar Catalog Database and $w_{50}^{off}$ from the FAST, The "off" state pulse width is constrained down to $w_{40}^{off}$, as $w_{30}^{off}$, $w_{20}^{off}$, and $w_{10}^{off}$ are impacted by off-pulse noise.

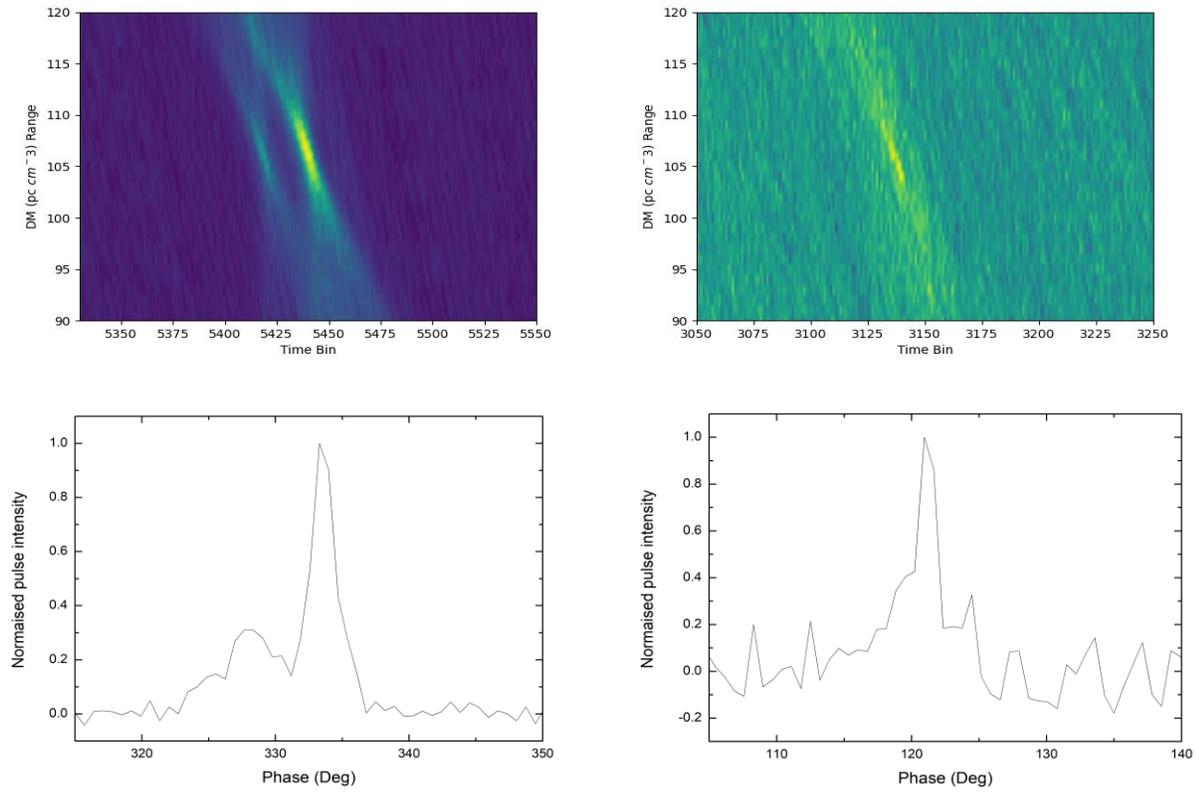

Fig. S1. DM VS time bin plot for "on" (upper left) and "off" (upper right) states, and the corresponding individual pulses are right under them with normalization. Time per bin equals 16 $\tau_s$.

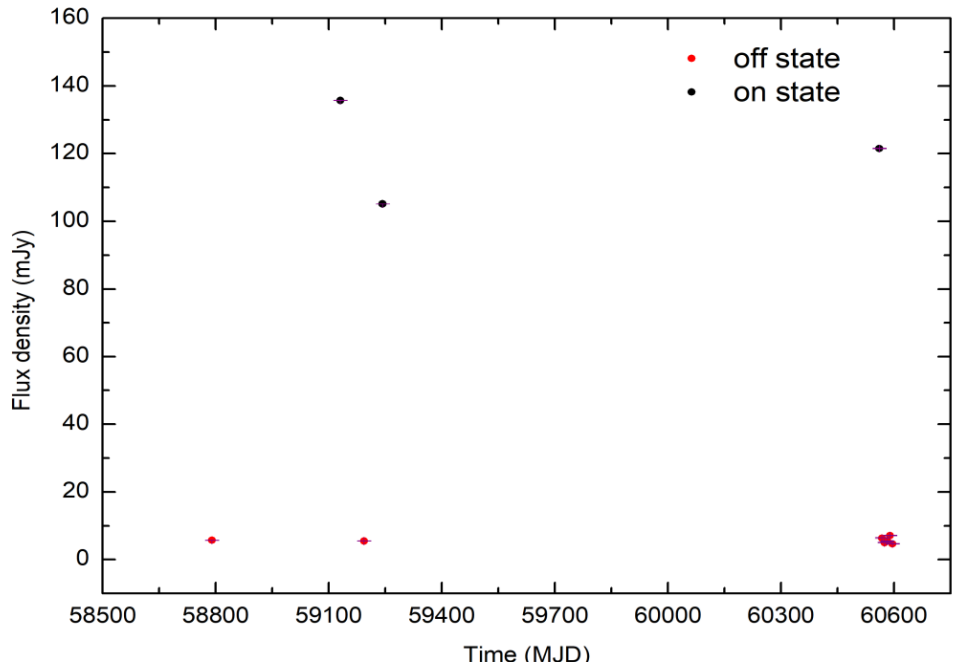

Fig. S2. The mean fluxes density across different observing epochs are shown. The red dots represent the mean flux density of the individual bursting dwarfs in the "off" states, while the black dots indicate the mean flux of all individual pulses in the "on" states.

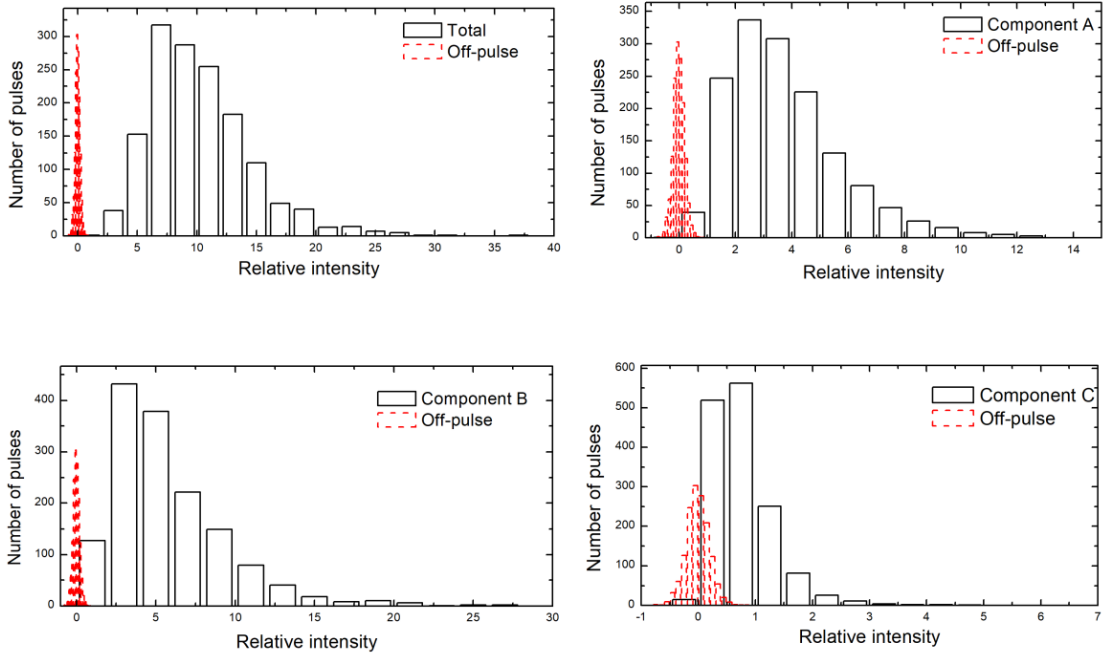
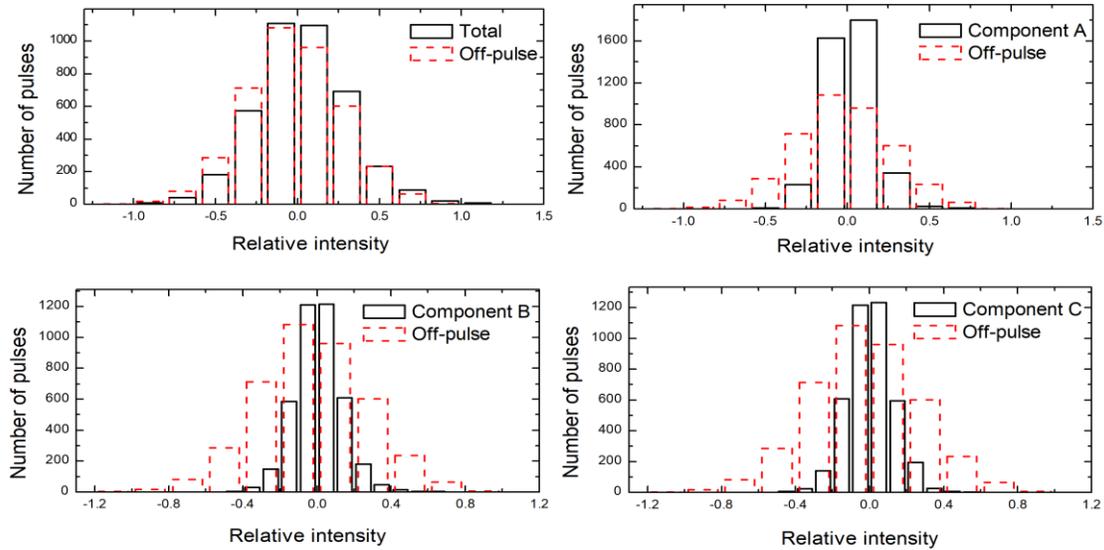

Fig. S3. Frequency count of intensity distribution of total, and its subcomponents for both "on" (top plot) and "off" (bottom plot) state. In each plot, the upper left panel displays the frequency count of the relative intensity distribution for both total and off-pulse noise. Similar plots, but with components A (upper right), B (lower left), and C (lower right) are arranged sequentially.

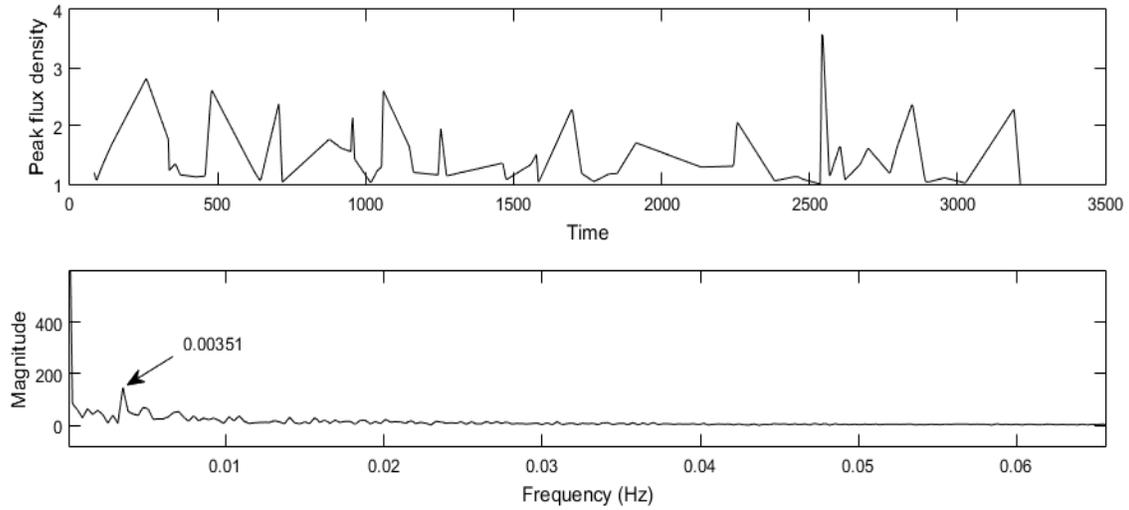

Fig. S4. The periodicity analysis of the individual bursting dwarfs of PSR B1931+24 in its "off" state (at 58790 MJD). The upper plot is peak flux of individual dwarf pulses vs time. The lower plot is fast Fourier transform to the upper plot data; the possible periodicity of the individual dwarfs is ~286 s, which is about ~351 rotation period of PSR B1931+24.

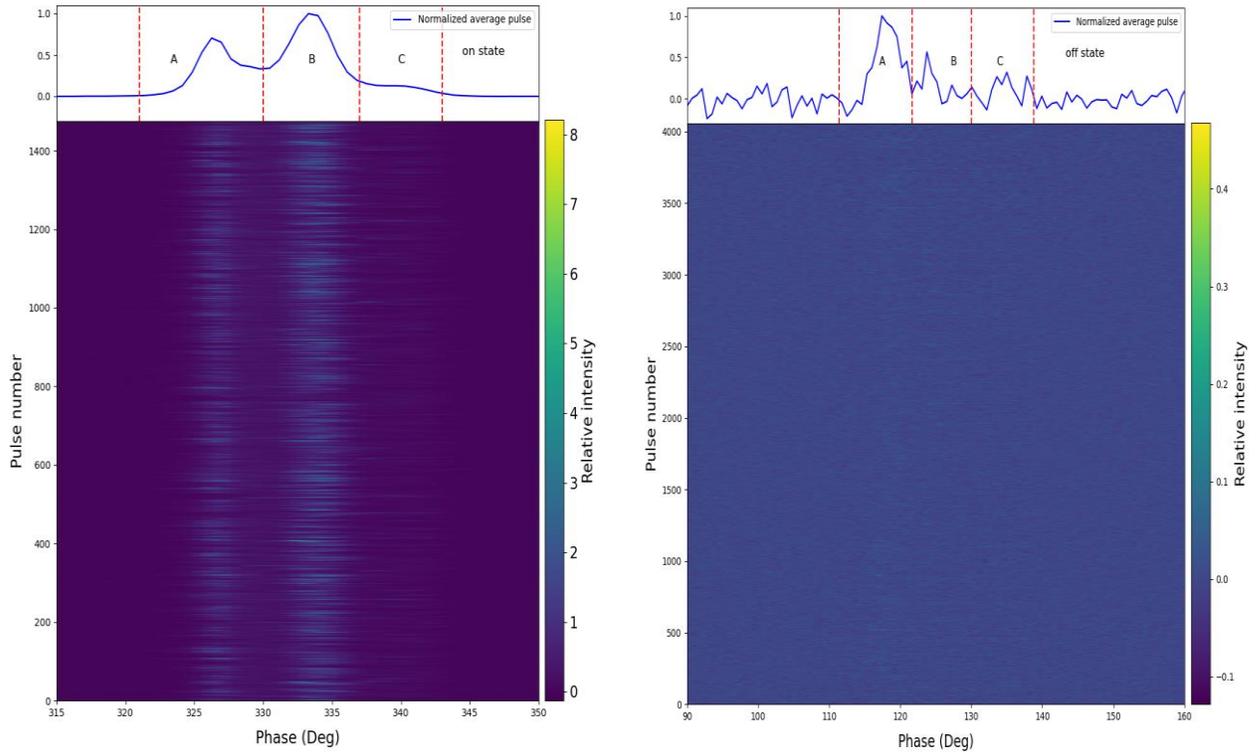

Fig. S5. The original pulse stacks of the "on" (left) and "off" state data at 59131 and 58790 MJDs, respectively.

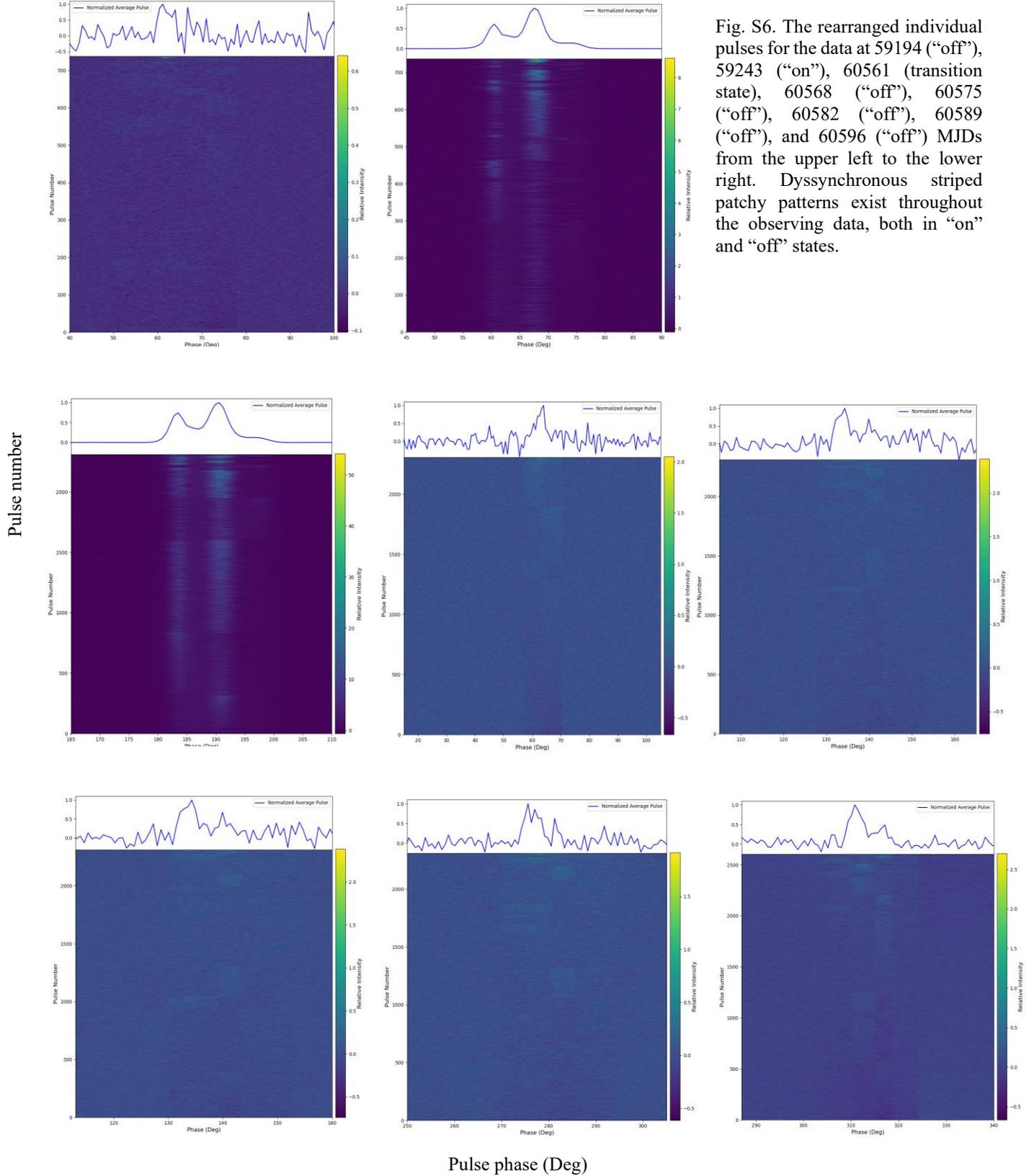

Fig. S6. The rearranged individual pulses for the data at 59194 ("off"), 59243 ("on"), 60561 (transition state), 60568 ("off"), 60575 ("off"), 60582 ("off"), 60589 ("off"), and 60596 ("off") MJDs from the upper left to the lower right. Dyssynchronous striped patchy patterns exist throughout the observing data, both in "on" and "off" states.

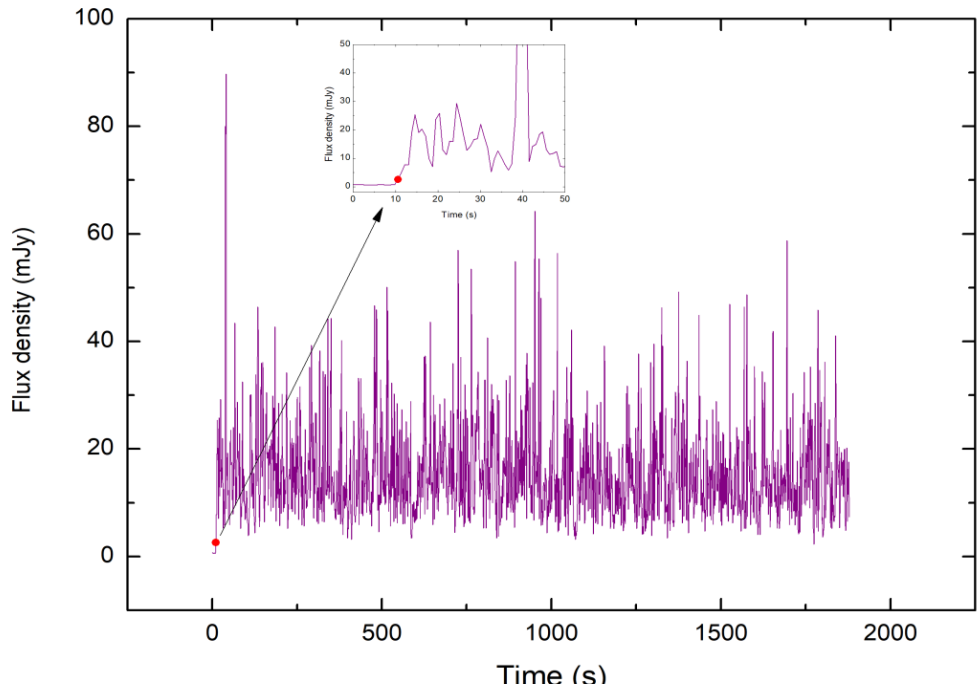

Fig. S7. Flux density evolution of pulse peak of the individual pulse at 60561 MJD. The red dot represents the peak flux density of the transitioning individual pulse (i.e., 14th) referred to in Fig. 4. It can be seen that the flux density is gradually increasing during the transition.